\documentclass[useAMS,usenatbib]{mn2e}

\usepackage{graphicx}
\usepackage{bm}

\title[Vertical convection in discs]{On the angular momentum transport due to vertical convection in accretion discs}
\author[G. Lesur and G. I. Ogilvie]{Geoffroy Lesur and Gordon I. Ogilvie\thanks{E-mail:
g.lesur@damtp.cam.ac.uk}\\
Department of Applied Mathematics and Theoretical Physics,\\
University of Cambridge, Wilberforce Road, Cambridge CB3 0WA, UK\\
}

\begin{document}

\date{Accepted 23/02/2010. Received ------------; in original form ---------------------}

\pagerange{\pageref{firstpage}--\pageref{lastpage}} \pubyear{2010}

\maketitle

\label{firstpage}

\begin{abstract}
The mechanism of angular momentum transport in accretion discs has long been debated. Although the magnetorotational instability appears to be a promising process, poorly ionized regions of accretion discs may not undergo this instability. In this letter, we revisit the possibility of transporting angular momentum by turbulent thermal convection.  Using high-resolution spectral methods, we show that strongly turbulent convection can drive outward angular momentum transport at a rate that is, under certain conditions, compatible with observations of discs. We find however that the angular momentum transport is always much weaker than the vertical heat transport. These results indicate that convection might be another way to explain global disc evolution, provided that a sufficiently unstable vertical temperature profile can be maintained.

\end{abstract}


\section{Introduction}
As shown by the early models of \cite{SS73} and \cite{LP74}, the structure and evolution of accretion discs are controlled by the process of
angular momentum transport. It is often assumed that turbulence is responsible for this transport, although its
nature and physical origin are barely understood. The most straightforward way to obtain a turbulent disc is to find a linear instability
of the basic flow, which could drive turbulence in the nonlinear regime. However, the presence of a turbulent
flow does not necessarily imply an outward transport of angular momentum through the disc, if the turbulence is sustained by sources of energy other than the Keplerian shear.

The magnetorotational instability (MRI), pointed out in the context of accretion discs by \cite{BH91a}, 
is believed to play an important role in accretion discs, as it produces turbulence that efficiently transports angular momentum outwards \citep[e.g.][]{B03}. 
However, the behaviour of the MRI in poorly ionized gas is uncertain. In some regions of discs, the instability
might cease to exist because the resistivity is too large, leading to the concept of `dead zones' \citep{G96}. On the other hand, 
purely hydrodynamical processes might also be at work in discs, such as the subcritical baroclinic instability \citep{KB03,PSJ07,LP10}, 
driven by radial entropy gradients. 

The presence of a vertical convective instability in discs was initially suggested by \cite{LP80} as a way of transporting angular momentum. They pointed out that the temperature-dependence of the opacity in protostellar (or protoplanetary) discs naturally leads to convective instability in certain regimes.
However, subsequent studies, using either linear approaches \citep{RG92}
or nonlinear simulations \citep{C96,SB96}, predicted a weak inward transport associated with convection, leading to the conclusion that this process was irrelevant for the problem of
accretion disc dynamics. Note that these simulations were performed in a `weakly nonlinear' regime with relatively low 
Reynolds and Rayleigh numbers \citep{C96}.

In this letter, we challenge these conclusions using high-resolution spectral methods and a simplified setup for the convection problem. In \S\ref{model}, we introduce
our model, the relevant dimensionless parameters and the numerical setup. We present simulations of convection with free-slip vertical boundary conditions in \S\ref{wall} and of
homogeneous convection in \S\ref{homogeneous}. Finally, we briefly discuss these results and their implications in \S\ref{discussion}.
 
\section{Model and equations}
\label{model}
\subsection{Model}
In the following, we consider a model flow in which rotation, shear and thermally induced convection are included. Instead of considering the full disc problem, with its particular density and temperature structure, we use the shearing box approximation \citep{HGB95,B03,RU08}. As a simplification, we assume the flow is incompressible and we introduce vertical stratification within the Boussinesq approximation \citep{SV60}. Although this approximation is not fully justified in a real disc, it allows us to explore the regime of strongly turbulent sheared convection in a more controlled way. In this letter we explore the way turbulent convection transports angular momentum, but we do not include the generation of an unstable temperature profile due to viscous heating. Instead, we impose a `background' stratification profile as prescribed by the Boussinesq approximation. Therefore, one has to keep in mind that convection is not generated self-consistently in our model.

The shearing-box equations are found by considering a Cartesian box centred at $r=R_0$, rotating with the disc at angular velocity $\Omega=\Omega(R_0)$. Defining $r-R_0 \rightarrow x$ and $R_0\phi \rightarrow y$ as in \cite{HGB95}, one obtains the following set of equations:
\begin{eqnarray}
\nonumber  \partial_t \bm{u}+\bm{u\cdot\nabla} \bm{u}&=&-\bm{\nabla} \Pi
-2\bm{\Omega \times u}\\
\label{motiongeneral}& &+2\Omega S x \,\bm{e_x}- \Lambda N^2\theta\, \bm{e_z}+\nu\Delta\bm{u},\\
\label{entropygeneral}\partial_t \theta +\bm{u\cdot\nabla}\theta&=&u_z/\Lambda+\chi\Delta\theta,\\
\label{divv} \bm{\nabla \cdot u}&=&0,
\end{eqnarray}
where $\bm{u}$ is the total velocity, $\theta\equiv \delta\rho/\rho$ is the perturbation of the density logarithm (or entropy, as pressure perturbations are much smaller in the Boussinesq approximation), $\nu$ is the kinematic viscosity and $\chi$ is the thermal diffusivity.
In these equations, we have defined the mean shear $S=-r\partial_r \Omega$, which is set to $S=(3/2)\Omega$ for a Keplerian disc. The generalised pressure $\Pi$ is calculated by solving a Poisson equation derived from the incompressibility condition. For dimensional consistency with the traditional Boussinesq approach, we have introduced a stratification length $\Lambda\equiv -g/N^2$, where $g$ is the vertical component of the gravity.  Note, however, that $\Lambda$ disappears from the dynamical properties of this system of equations as one can renormalize the variables by defining $\theta'\equiv\Lambda \theta$.
The stratification itself is controlled by the Brunt--V\"ais\"al\"a frequency $N$, defined for a perfect gas by
\begin{equation}
N^2=-\frac{1}{\gamma \rho}\frac{\partial P}{\partial z}\frac{\partial}{\partial z}\ln\Big(\frac{P}{\rho^\gamma}\Big),
\end{equation}
where $P$ and $\rho$ are the pressure and density of the background equilibrium profile and $\gamma$ is the adiabatic index. Physically, $N$ corresponds to the oscillation frequency of a fluid particle displaced vertically from its equilibrium position. When the flow is convectively unstable, one has $N^2<0$. In our model, we assume $N$ is constant, which formally corresponds to a local model in $z$.

One can easily check that the velocity field $\bm{U}=-Sx\bm{e_y}$ is a steady solution of equations (\ref{motiongeneral})--(\ref{entropygeneral}). In the following we consider the evolution of the perturbations $\bm{v}$ (not necessarily small) of this profile defined by $\bm{v}=\bm{u}-\bm{U}$.

In the following, we use shearing-sheet boundary conditions \citep{HGB95} in the $x$ direction and periodic boundary conditions in $y$. In the $z$ direction, we use either free-slip boundary conditions (\S\ref{wall}), imposing $v_z=\theta=0$ and $\partial_z v_x=\partial_z v_y=0$, or periodic boundary conditions (\S\ref{homogeneous}). The free-slip vertical boundary conditions correspond to a rigid wall with no tangential stress and a fixed temperature. Similar boundary conditions for the velocity field were used by \cite{C96}.

\subsection{Dimensionless numbers}
To simplify the analysis of convection in sheared flow, we will use three dimensionless numbers defined as follows:
\begin{itemize}

\item The Richardson number $Ri=-N^2/S^2$ compares buoyancy forces to the mean shear. In a real disc, this number is determined by the balance between viscous heating and vertical heat transport. Assuming a typical velocity $v$ due to convective motions, we define the turbulent viscosity using mixing length theory $\nu_t\sim \delta v H$, where $\delta$ is an efficiency factor and $H$ is the disc thickness. The turbulent heat transport is on the other hand $F\sim \rho v^3$. Assuming thermal balance, we get $\rho v^3/H \sim  \rho \nu_t \Omega^2$ leading to $v^2\sim \delta c_s^2$ where $c_s$ is the local sound speed. Since turbulence is driven by the unstable entropy gradient, we have in first approximation $v\sim |N|H$, so that $Ri\sim \delta^{1/2}$. As we will see in the following, we generally get $\delta \ll 1$, meaning that the regime $Ri<1$ is relevant to accretion discs.

\item The Prandtl number $Pr=\nu/\chi$ compares viscous dissipation to thermal diffusion. Using standard values for the viscosity and thermal conductivity of $H_2$, relevant to a protoplanetary disc, one has
\begin{eqnarray}
\nonumber Pr=1.5\times 10^{-9}\, \Big(\frac{T}{300\,\mathrm{K}}\Big)^{-5/2}\Big(\frac{\kappa}{1\,\mathrm{cm}^2/\mathrm{g}}\Big)\Big(\frac{\rho}{10^{-9} \, \mathrm{g/cm}^3}\Big),
\end{eqnarray}
where $\kappa$ is the opacity.
\item The Rayleigh number $Ra=-N^2L^4/\chi\nu$ compares buoyancy forces to dissipation processes. In a disc, one gets
\begin{eqnarray}
\nonumber \frac{Ra}{Ri}&=&9\times 10^{20} \, \Big(\frac{L}{R}\Big)^4\Big(\frac{r}{\mathrm{AU}}\Big )\Big(\frac{\kappa}{1\,\mathrm{cm}^2/\mathrm{g}}\Big)\Big(\frac{M}{M_\odot}\Big) \times\\
\nonumber & &\Big(\frac{T}{300\,\mathrm{K}}\Big)^{-7/2} \Big(\frac{\rho}{10^{-9} \, \mathrm{g/cm}^3}\Big)^3.
\end{eqnarray}

\end{itemize}
In the following, we will assume $Pr=1$ for simplicity. A more complete study including the effect of varying $Pr$ will be the subject of a subsequent paper. The length $L$ introduced in these definitions is assumed to be a typical dimension of the box and should be of the order of the disc thickness. In the following, we define $L$ as the smallest box length with $L=L_z$ in \S\ref{wall} and $L=L_x$ in \S\ref{homogeneous}. When not explicitly mentioned, the unit of time is the shear timescale $S^{-1}$ which is equal to $(3\pi)^{-1}$ orbital periods. In the following, all of our simulations are computed in the shear dominated regime ($Ri<1$) which is expected in discs.

\subsection{Turbulent transport and energy budget} 
By analogy with \cite{SB96}, we define the dimensionless transport of angular momentum through the $\alpha$-like coefficient $\alpha \equiv \langle v_x v_y \rangle / S^2L^2$, $\langle \cdot \rangle$ being a time and space average. Since $S=1$ and $L=1$ in all the simulations presented below, we will simply write $\alpha=\langle v_x v_y \rangle$ in our units. 

Multiplying the $y$ component of (\ref{motiongeneral}) by $v_y$, integrating over the volume of the box and using the boundary conditions presented above, we obtain the energy budget of the fluctuations in the $y$ direction,
\begin{equation}
\label{Eq:Ebudget}
\partial_t \overline{v_y^2}=\overline{\Pi\partial_y v_y}+(S-2\Omega)\overline{v_x v_y}-\nu \overline{|\bm{\nabla} v_y|^2},
\end{equation}
where $\overline{X}$ denotes the volume average of $X$. Considering a quasi-steady turbulence and assuming the pressure--strain correlation $\overline{\Pi\partial_y v_y}$ to be negligible, \cite{SB96} concluded from this equation that $(S-2\Omega)\overline{v_x v_y}>0$, leading to $\alpha<0$ in a disc. This hypothesis was motivated by numerical simulations \citep{SB96,C96} in which the turbulence was weakly non-axisymmetric, leading to the conclusion that $y$ derivatives should be negligible. It is however well known that the pressure--strain correlation tensor is responsible for the redistribution of energy in fully developed anisotropic turbulence \citep{S91,P00}. Therefore, when the flow becomes strongly nonlinear (i.e. at large $Ra$), there is no reason to assume that this term is negligible.

\subsection{Numerical model}

We use the Snoopy code to solve equations (\ref{motiongeneral})--(\ref{divv}). This code uses a 3D Fourier representation of the flow with a third-order timestepping scheme to have an excellent accuracy at all scales. It has been used in several studies of hydrodynamic and MHD turbulence \citep[see e.g.][]{LL05,LL07}. When working with free-slip boundary conditions, a sine--cosine decomposition is used instead of the Fourier basis in the $z$ direction \citep[see e.g.][]{CEW03}.
 
\section{Convection with walls}
\label{wall}

\begin{figure}
   \centering
   \includegraphics[width=0.9\linewidth]{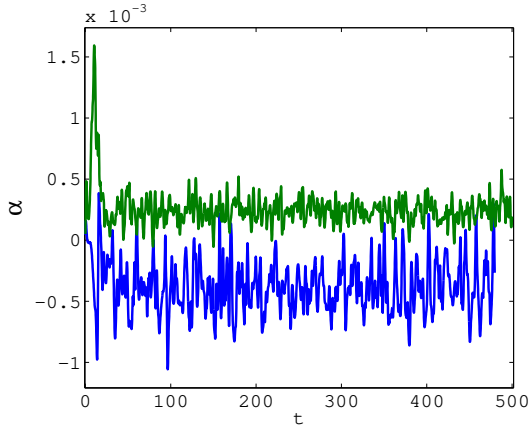}
   \caption{Turbulent transport as a function of time for run W1 (lower curve) and W8 (upper curve). The modification of the average turbulent transport is due to the Rayleigh number (see text).}
              \label{Tserie}%
\end{figure}

In the first set of simulations, we introduce free-slip boundary conditions in the $z$ direction. These boundary conditions prevent the development of large-scale vertical flows which are unrealistic for a disc (see \S\ref{homogeneous}). 
In the particular setup considered here, convection starts at $Ra\simeq 4100$ for $Ri=0.4$ and $Ra\simeq 6900$ for $Ri=0.2$. 

In order to study the dependence of the turbulent transport of angular momentum on our dimensionless parameters, we have carried out several simulations which are presented in Tab.~\ref{FStable}. Average values are computed from simulations run for $500\,S^{-1}$, excluding the first $100\,S^{-1}$ from the averaging procedure to eliminate the influence of the initial conditions. As shown in Fig.~\ref{Tserie}, the simulations are sufficiently homogeneous on these timescales for this average to be meaningful.

\subsection{Rayleigh number and transport}
\begin{table*}
\begin{tabular}{lllrrrrr}
\hline
Run & Resolution & Box size & $Ra$ & $Ri$ & $\langle \frac{1}{2}\bm{v}^2\rangle/S^2L^2 $ & $\langle v_x v_y \rangle/S^2L^2\equiv \alpha $  & $\Lambda \langle \theta v_z \rangle/SL^2 $\\
       & $n_x\times n_y \times n_z$& $L_x\times L_y \times L_z$ & & & & & \\
\hline
W1 & $256\times 256 \times 128$ & $2\times 4 \times 1$ & $1\times 10^5$ & 0.4 & $5.7\times 10^{-3}$ &$-3.9\times 10^{-4}$ & $7.7\times 10^{-3}$\\
W2 & $256\times 256 \times 128$ & $2\times 4 \times 1$ & $3\times10^5$ & 0.4 & $7.0\times 10^{-3}$ & $-2.3\times 10^{-4}$ & $7.4\times 10^{-3}$\\
W3 & $256\times 256 \times 128$ & $2\times 4 \times 1$ & $6\times10^5$ & 0.4 & $6.9\times 10^{-3}$ & $-1.4\times 10^{-4}$ & $7.0\times 10^{-3}$\\
W4 & $256\times 256 \times 128$ & $2\times 4 \times 1$ & $1.2\times10^6$ & 0.4 & $7.2\times 10^{-3}$ & $-4.6\times 10^{-5}$ & $6.2\times 10^{-3}$\\
W5 & $256\times 256 \times 128$ & $2\times 4 \times 1$ & $2.5\times10^6$ & 0.4 & $7.2\times 10^{-3}$ & $5.8\times 10^{-5}$ & $5.7\times 10^{-3}$\\
W6 & $256\times 256 \times 128$ & $2\times 4 \times 1$ & $5\times10^6$ & 0.4 & $7.1\times 10^{-3}$ & $1.4\times 10^{-4}$ & $5.1\times 10^{-3}$\\
W7 & $256\times 256 \times 128$ & $2\times 4 \times 1$ & $1\times10^7$ & 0.4 & $6.7\times 10^{-3}$ & $2.1\times 10^{-4}$ & $4.6\times 10^{-3}$\\
W8 & $256\times 256 \times 128$ & $2\times 4 \times 1$ & $1.4\times10^7$ & 0.4 & $6.4\times 10^{-3}$ & $2.4\times 10^{-4}$ & $4.3\times 10^{-3}$ \\
\hline
W9   & $256\times 256 \times 128$ & $2\times 4 \times 1$ & $   3\times10^5 $ & 0.2 & $ 2.1\times 10^{-3}$ & $-1.6\times 10^{-4}$ & $3.9\times 10^{-3}$\\
W10 & $256\times 256 \times 128$ & $2\times 4 \times 1$ & $1.2\times10^6 $ & 0.2 & $ 2.2\times 10^{-3}$ & $-1.0\times 10^{-4}$  & $3.7\times 10^{-3}$\\
W11 & $256\times 256 \times 128$ & $2\times 4 \times 1$ & $ 5 \times10^6 $ & 0.2 & $ 2.2\times 10^{-3}$ & $-5.3\times 10^{-5}$  & $3.1\times 10^{-3}$\\
W12 & $256\times 256 \times 128$ & $2\times 4 \times 1$ & $ 2 \times10^7 $ & 0.2 & $ 2.0\times 10^{-3}$ & $-8.7\times 10^{-6}$  & $2.5\times 10^{-3}$\\
W13 & $512\times 512 \times 256$ & $2\times 4 \times 1$ & $ 8 \times10^7 $ & 0.2 & $ 1.8 \times 10^{-3}$ & $3.0\times 10^{-5}$  & $1.9\times 10^{-3}$ \\
W14 & $512\times 512 \times 256$ & $2\times 4 \times 1$ & $ 3.2 \times10^8 $ & 0.2 & $ 1.7 \times 10^{-3}$ & $5.8\times 10^{-5}$ & $1.6\times 10^{-3}$ \\
\hline
\end{tabular}
\caption{\label{FStable}List of simulations with free-slip boundary conditions.  The turbulent kinetic energy ($6^\mathrm{th}$ column), angular momentum transport ($7^\mathrm{th}$ column) and heat transport ($8^\mathrm{th}$ column) are computed from simulations.}
\end{table*}

\begin{figure}
   \centering
   \includegraphics[width=0.8\linewidth]{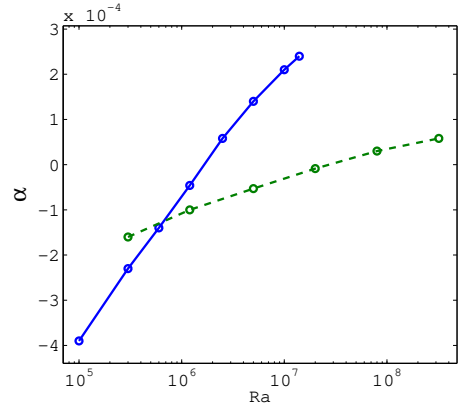}
   \caption{Correlation between the turbulent transport and $Ra$ for $Ri=0.4$ (plain line) and $Ri=0.2$ (dashed line).}
              \label{WallCorr}%
\end{figure}

The most striking observation in these simulations is the correlation between $Ra$ and the turbulent transport $\alpha$. In particular, for $Ra<10^6$ we recover the result of \cite{SB96} of a small and negative transport (runs W1--W4 and W9--W12). However, for sufficiently large $Ra$, the sign of the transport is reversed. To demonstrate this effect, we show in Fig.~\ref{WallCorr} the dependence of the turbulent transport on $Ra$, for $Ri=0.4$ and $Ri=0.2$.

It is possible to deduce a general scaling for the transport from these results:
\begin{equation}
\alpha\simeq -1.8\times 10^{-3}+2.9\times 10^{-4} \log_{10}(Ra)
\end{equation}
for $Ri=0.4$ and
\begin{equation}
\alpha\simeq -5.4\times 10^{-4}+7.2\times 10^{-5}\log_{10}(Ra)
\end{equation}
for $Ri=0.2$. Surprisingly, these scalings predict a transport of the order of $10^{-3}$ for values of $Ra$ which could be expected in a disc.
However, this conclusion is challenged by the fact that at large $Ra$, the dependence of $\alpha$ on $Ra$ should become weaker since the length-scales associated with transport are then much larger than the diffusive scales and the diffusion coefficients may be expected to become less important. A weak effect of this kind is already observed in Fig.~\ref{WallCorr} for $Ra>10^7$.

Interestingly, changing the Rayleigh number does not significantly modify the turbulent kinetic energy of the system. This clearly demonstrates that the correlation observed above is \emph{not} due to a modification of the turbulent activity in the flow but results instead from a change in the anisotropy of the turbulence.  Note also that the turbulent heat transport $\langle \theta v_z\rangle $ decreases as $Ra$ increases. However, we always find $\langle \theta v_z\rangle \gg \langle v_xv_y\rangle $, showing that convection is inefficient at transporting angular momentum. This point will be examined further in \S\ref{discussion}.

\subsection{Energy budget}
\begin{figure}
   \centering
   \includegraphics[width=0.9\linewidth]{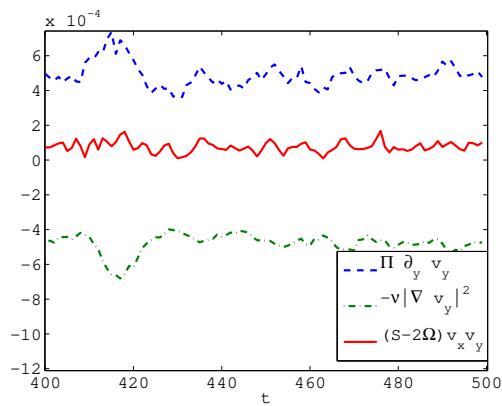}
   \caption{Energy budget in the $y$ direction (eq.~\ref{Eq:Ebudget}) from run W7. }
              \label{Fig:Ebudget}%
\end{figure}
\begin{figure}
   \centering
   \includegraphics[width=0.9\linewidth]{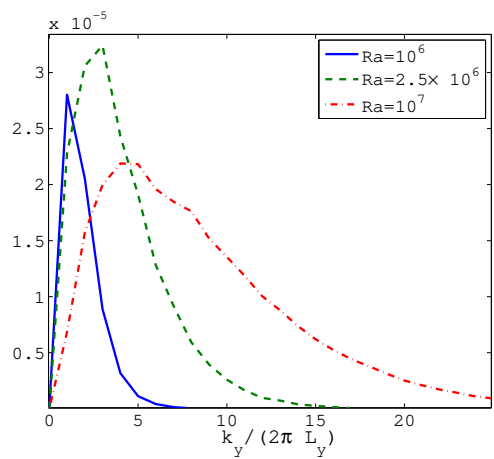}
   \caption{1D spectra of the pressure--strain correlation from run W1 (plain line), W5 (dashed line) and W7 (dash-dotted line).}
              \label{Fig:spectrum}%
\end{figure}

To check if the pressure--strain correlation is important in our simulations, we plot in Fig.~\ref{Fig:Ebudget} the time history of each term of equation~(\ref{Eq:Ebudget}) for run W7. As shown in this plot, the energy balance at high $Ra$ is dominated by an equilibrium between the pressure--strain correlation and the dissipation term. Moreover, the turbulent transport term is \emph{negligible} in this situation. This clearly indicates that at high enough $Ra$, the hypothesis of \cite{SB96} is no longer valid and the turbulent transport of angular momentum can be directed outward.

The presence of a large pressure--strain correlation term in equation~(\ref{Eq:Ebudget}) shows that non-axisymmetric structures are playing an important role at large $Ra$. To isolate the role played by these non-axisymmetric structures, we plot in fig.~\ref{Fig:spectrum} the 1D spectrum of the pressure--strain correlation. As expected, we find no contribution from axisymmetric modes ($k_y=0$). However, as $Ra$ becomes larger, the contribution from smaller-scale non-axisymmetric modes becomes more important. This effect is evident in run W7 where contributions from scales as small as $L_y/20$ are still important.

This result clearly indicates that as the flow becomes turbulent, it develops small-scale motions which tend to redistribute the energy fluctuations in the three directions. This is consistent with the picture of anisotropic turbulence becoming isotropic at small scales thanks to the pressure--strain correlation. It also indicates that the results of \cite{C96} and \cite{SB96} were due to a too small $Ra$, an issue already mentioned by \cite{C96}.

\section{Homogeneous convection}
\label{homogeneous}
\begin{table*}
\begin{tabular}{lllrrrrr}
\hline
Run & Resolution & Box size & $Ra$ & $Ri$ & $\langle \frac{1}{2}\bm{v}^2\rangle/S^2L^2 $ & $\langle v_x v_y \rangle/S^2L^2\equiv \alpha $  & $\Lambda \langle \theta v_z \rangle/SL^2 $\\

       & $n_x\times n_y \times n_z$& $L_x\times L_y \times L_z$ & & & & & \\
\hline
H1 & $64\times 64 \times 128$ & $1\times 1 \times 2$ & $10^4$ & 0.1 & $2.8\times 10^{-2}$ & $-3.4\times10^{-4}$ & $1.1\times 10^{-1}$\\
H2 & $64\times 64 \times 128$ & $1\times 1 \times 2$ & $4\times 10^4$ & 0.1 & $1.8\times 10^{-2}$ & $9.6\times10^{-4}$& $6.2\times 10^{-2}$ \\
H3 & $128\times 128 \times 256$ & $1\times 1 \times 2$ & $4\times 10^5$ & 0.1 & $1.6\times 10^{-2}$ & $9.0\times10^{-4}$ & $4.2\times 10^{-2}$\\
H4 & $128\times 128 \times 256$ & $1\times 1 \times 2$ & $4\times 10^6$ & 0.1 & $1.9\times 10^{-2}$ & $1.6\times10^{-3}$ & $3.7\times 10^{-2}$\\
H5 & $128\times 256 \times 256$ & $1\times 2 \times 2$ & $4\times 10^5$ & 0.1 & $1.5\times 10^{-2}$ & $5.3\times10^{-4}$ & $4.0\times 10^{-2}$\\
H6 & $128\times 256 \times 256$ & $1\times 2 \times 2$ & $4\times 10^6$ & 0.1 & $1.3\times 10^{-2}$ & $9.0\times10^{-4}$ & $2.8\times 10^{-2}$\\
\hline
\end{tabular}
\caption{\label{Htable}List of simulations with periodic boundary conditions in the $z$ direction. }
\end{table*}

To test the possible limitations of the free-slip boundary conditions used above, we have also considered the problem of homogeneous convection, using periodic boundary conditions in the vertical direction applied to $\bm{v}$ and $\theta$ \citep{BO97}. In this context, the flow becomes convectively unstable for $Ra>1558$, developing first ``elevator'' modes with a purely vertical motion $v_z\propto \cos(k_x x+\phi)$. Since these modes are also exact nonlinear solutions of equations (\ref{motiongeneral})--(\ref{divv}), they can grow exponentially for long times, leading to ``spikes'' in the turbulent energy and transport. These exponential growth phases are unrealistic, as they should be broken by secondary instabilities or the effects of density stratification, which are neglected in the Boussinesq approximation. A way to limit the formation and amplification of these modes is to consider ``tall'' boxes with $L_z>L_x$, so that secondary instabilities are more efficient at breaking up the elevator modes. Note that the phenomenology of these elevator modes is very similar to that of channel mode solutions found in the context of the magnetorotational instability \citep[see][]{GX94}.

\begin{figure}
   \centering
   \includegraphics[width=0.9\linewidth]{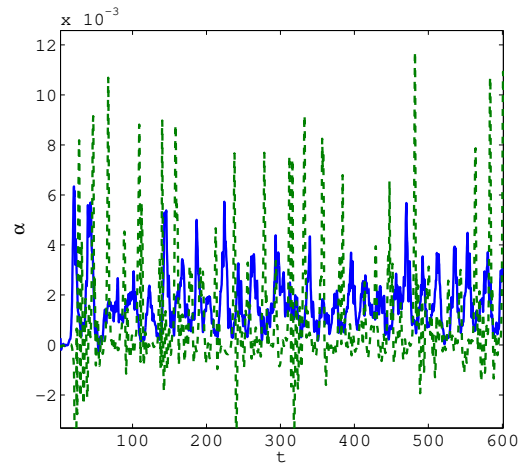}
   \caption{Turbulent transport as a function of time for run H2 (dashed line) and H4 (plain line).}
              \label{Fig:TserieH}%
\end{figure}

We have used this tall box setup with several values of $Ra$ and $L_y$ in the regime of weak convection, $Ri=0.1$ (Tab.~\ref{Htable}). Looking at the time series (Fig.~\ref{Fig:TserieH}), we note the presence of large peaks of transport at small $Ra$, with both positive and negative values. These peaks are associated with the growth of elevator modes to high amplitude and their break-up through secondary instabilities. We note however that this behaviour disappears at larger $Ra$, although large fluctuations remain, leading to a positive-definite turbulent transport with an average value $\alpha\sim 10^{-3}$.

Modifying the size of the box in the $y$ direction does not change dramatically the results, and on average, we find a transport that is stronger in the homogeneous case than with free-slip boundary conditions. Note also that the average turbulent kinetic energy is larger in homogeneous runs (Tab.~\ref{Htable}) which partly explains the higher turbulent transport observed in those cases. 

\section{Discussion\label{discussion}}
\label{discussion}
In this letter, we have shown that turbulent convection, if present in accretion discs, could lead to an \emph{outward} transport of angular momentum, provided that the Rayleigh number is large enough (i.e.\ molecular viscosity and thermal diffusivity are small enough). We also conjectured that the previous negative results \citep{C96,SB96}  were due to a too small Rayleigh number, leading to nearly axisymmetric turbulence which cannot transport angular momentum outward.

Despite this positive result, this letter does not demonstrate that convection is actually present in discs and is responsible for the observed accretion rates. To demonstrate this point, one probably has to include viscous heating and a realistic model of disc stratification, which are beyond the scope of this letter. 

We note that the use of the Boussinesq approximation with constant stratification is a strong simplification of the problem. However, we always find $v<0.1SL$, showing that our incompressible approximation is valid. Moreover, we believe that the Rayleigh number effect found in this work will remain valid with a more realistic vertical profile, as small scales are weakly influenced by the global (vertical) structure.

Interestingly, we find that the turbulent transport of angular momentum is always much weaker than the turbulent heat transport (typically $\langle \theta v_z \rangle \sim 20\, \alpha$).  
This is to be expected as turbulent convection is not shear driven, but it also indicates that convection is inefficient at transporting angular momentum. This leads us to conjecture that it might not be possible to generate vertical convection self-consistently with viscous heating \emph{only}. Instead, one might have to consider an additional heating source which could be due to radiation or chemical reactions.

\section*{Acknowledgments}
GL thanks Steven Balbus and John Papaloizou for useful and stimulating discussions. Some of the simulations presented in this paper were performed using the Darwin Supercomputer of the University of Cambridge. The authors acknowledge support by STFC. 

\bibliographystyle{mn2e}
\bibliography{glesur}

\label{lastpage}

\end{document}